\documentstyle[12pt,aasms4]{article}

\lefthead{Dolphin}
\righthead{WLM Star Formation History}

\begin{document}

\title{HST Studies of the WLM Galaxy. II. The Star Formation History from Field Stars
\footnote{
Based on observations with the NASA/ESA Hubble Space Telescope obtained
at the Space Telescope Science Institute, which is operated by the
Association of Universities for Research in Astronomy, Inc., under
NASA contract NAS5-26555.
}}

\author{Andrew E.\ Dolphin}
\affil{National Optical Astronomy Observatories, PO Box 26732, Tucson, AZ 85726 \\
dolphin@noao.edu}

\begin{abstract}
HST F555W and F814W photometry of a portion of the WLM galaxy are presented.  The distance modulus is determined via fitting of the entire color-magnitude diagram to be $(m-M)_0 = 24.88 \pm 0.09$, which is consistent with the RGB tip distance.  The galaxy's measurable star formation history appears to have begun no more than 12 Gyr ago, with about half of WLM's total star formation (by mass) formed before 9 Gyr ago.  The star formation rate gradually decreased, until a recent increase in activity starting between 1 and 2.5 Gyr ago.  This is still continuing to the present time, and is concentrated in the bar of the galaxy, as shown by the difference in recent star formation rates in the three WF chips.
\end{abstract}


\section{Introduction}

The WLM galaxy (DDO 221) is a Local Group dwarf irregular (class Irr IV-V), in an isolated part of the Local Group approximately 900 kpc from the Milky Way.  It was discovered by Wolf (1909) and rediscovered by Lundmark and Melotte (Melotte 1926), thus earning its name of Wolf-Lundmark-Melotte, abbreviated WLM.  Its distance is approximately 900 kpc, making it close enough to allow ground-based studies of its red giant branch (RGB) tip ($V = \sim22.5$) and upper main sequence, but distant enough that such studies will not reach the horizontal branch ($V = \sim25.4$).

Ables \& Ables (1977) used electrographic techniques to make photometric measurements of supergiants in WLM, getting $B$ and $V$ photometry for 50 stars, the faintest having a magnitude $B$ of 21.6.  With this photometry, they were able to estimate a distance modulus of $(m-M) = 26.23 \pm 0.14$.  They also noted a ``slightly diffuse starlike object about 2' west of the center'' of the galaxy, which they determined was most likely a globular cluster with $V=16.56$ and $(B-V)=0.67$.
Finally, they made contour maps in $B$ and $V$ of the galaxy, finding a color gradient with $(B-V)=0.26$ in the core and $(B-V)=0.95$ in the outer regions.  Sandage \& Carlson (1985) studied the brightest supergiants in WLM using a photometer and determined periods for 15 Cepheids.  With the Cepheid results, the distance modulus was heavily revised to $(m-M) = 24.93$, a drop of 1.3 magnitudes.

The first CCD study of this galaxy was made by Ferraro et al. (1989), with later work made by Minniti \& Zijlstra (1997).  Both works reached $V = \sim24$, deep enough to accurately measure the RGB tip but not the HB.  The Ferraro et al. (1989) data consisted of two fields, both partially in the bar and partially in the halo.
They were able to conclude that the two fields had different star formation rates in the very recent past (and therefore that WLM's star formation was localized rather than global), and that a Salpeter IMF was adequate for their data.

Minniti \& Zijlstra (1997) made a $V$ and $I$ study of WLM, which allowed them to study the older stars in greater detail.  They found that star formation began at least 10 Gyr ago in the galaxy, with a metallicity [Fe/H] of -1.45 $\pm$ 0.2.  A distance modulus of $(m-M)_0 = 24.75 \pm 0.1$ was calculated from the RGB tip.  Importantly, they observed that there was a lack of bright main sequence stars in the galaxy's halo, and therefore the color gradient observed by Ables \& Ables (1977) was due to a stellar population gradient.

With HST observations of WLM, we are now able to peer much deeper into the stellar content of this galaxy, with quality $V$ photometry to magnitude 27 ($M_V=\sim+2$).  A distance modulus of $(m-M)_0 = 24.73 \pm 0.07$ was calculated for the globular cluster from color-magnitude diagram fitting (Hodge et al. 1999), which was centered in the PC camera of WFPC2.  The data from the WF cameras also provide the opportunity to study the star formation history in a quantitative manner, using the method detailed in Dolphin (1997).

The metallicity enrichment history has not been studied extensively for this galaxy either, with only two solid estimates of the metallicity.  Hodge \& Miller (1995) used spectra of two HII regions in WLM to deduce a current metallicity of approximately -1.15 $\pm$ 0.2, while Minniti \& Zijlstra (1997) estimated that the old stars have a metallicity of -1.45 $\pm$ 0.2 through photometry.  These two figures provide rough endpoints of the enrichment history, but are very uncertain (to the point that both are consistent with -1.3).  A better understanding of this galaxy's enrichment history is possible through the HST data presented here.

Goals for this study will be:
\begin{itemize}
\item Determining the distance from CMD analysis, and comparing it with that obtained by Hodge et al. (1999) and an RGB tip analysis.
\item Dating the oldest stars in the galaxy, and determination of star formation rates since that time.
\item Determining the chemical enrichment history and comparing with estimates of the old star metallicity and measurements of the HII region metallicities.
\item Determining the nature of the stellar population differences. Has the star formation rate in the bar increased recently, that in the halo decreased, or a combination of the two?
\item Quantifying the recent star formation history differences: how much different are the recent star formation rates in different regions, and how long have they been different?
\end{itemize}

\section{Data}

\subsection{Observations}

As part of a Cycle 6 HST GO program, four images of part of WLM were taken on 28 September, 1998.  Two images were taken through the F555W filter with exposure times of 2600 and 2700 seconds, and two were taken through the F814W filter with exposure times of 2700 seconds each.  The PC chip was centered on WLM's globular cluster, the data for which was analyzed separately (Hodge et al. 1999), and the orientation of the camera was such that the WF chips lay approximately along the galaxy's minor axis, providing a sample of WLM bar and halo stars.
The alignment of the chips put WF2 the closest to the bar of the galaxy, WF3 further out, and WF4 well outside the bar.  Cosmic ray cleaning was made with a routine similar to that of the IRAF task CRREJ.  The cleaned $V$ image is shown in Figure \ref{fig-image}.

\placefigure{fig-image}

\subsection{Reduction}

Stellar photometry was carried out using the HSTPhot package (Dolphin 1999) used in Hodge et al. 1999.  This program is optimized for undersampled WFPC2 images, and uses a hybrid PSF fitting and aperture photometry routine to determine the position and brightness of each star located.  CMDs are given for all three WF chips combined, and for each chip separately, and are shown in Figure \ref{fig-cmds}.

\placefigure{fig-cmds}

\section{Analysis}

\subsection{Star Formation History Solutions}

For all CMD analyses in this paper, the Padova isochrones (Girardi et al. 1996, Fagotto et al. 1994) were used for the generation of synthetic CMDs.  These models include later (post-helium flash) phases of stellar evolution, despite the large uncertainties in those phases, providing completeness at the cost of accuracy.  For the purpose of matching the entire observed CMD with synthetic data, however, it is far better to use models with slightly incorrect AGB, red clump, and horizontal branch phases than models that omit them entirely.

  The star formation history solution method used here is based on that described in Dolphin (1997), and involves modeling the observed CMD with synthetic CMDs.  This process involves three steps.  First is the generation of synthetic CMDs, which requires theoretical isochrones, an artificial star library, and parameters of star formation: star formation rate, metallicity (both functions of time), distance, IMF, and extinction.  For this data set, which is not deep enough to observe below the ancient main sequence turnoff, a Salpeter IMF was assumed, with endpoints of 0.1 and 120 $M_\odot$.  For a single solution, a distance, extinction, and metallicity history would be assumed, while the star formation rates would be the free parameters.  Since any complex population (multiple ages and abundances) can be modeled as the sum of its component populations, its CMD is thus the sum of the CMDs of the component populations.  Thus all that is required to generate any number of CMDs with various past star formation rates is a set of CMDs of each of the individual component populations.  This was accomplished by dividing the time from the beginning of the galaxy's star formation to the present into segments, each with a constant star formation rate of 1 $M_\odot$/yr.  By multiplying each CMD by the average star formation rate over its period, one will then have the total synthetic CMD.

The second step is the conversion of the data into a format that can be fit.  To accomplish this, each CMD was divided into a large number of square bins, with the number of stars falling in each counted.  The synthetic CMDs, generated from the theoretical isochrones and the artificial star library, were given the same treatment.  Two sizes of bins were used for fits.  A lower-resolution fit, with bin size 0.1 mag in $V-I$ and 0.3 in $V$, would be more sensitive to larger features, such as luminosity functions along the main sequence and RGB.  A higher-resolution fit, with bin sizes half those of the lower-resolution fit, would be more sensitive to details.

Finally, a fit must be made for the star formation rates of each period of time, attempting to find the star formation rates for the single population CMDs that, when the CMDs are combined, will create the ``best'' fit to the observed data.  A standard least-$\chi^2$ fit can be easily thrown off when dealing with common observational errors, most notably the presence of observed ``stars'' (cosmic rays, noise, binary stars, Galactic foreground contamination, etc) in regions of the CMD where the models do not predict any stars.  Such a case will give a number of expected stars of $0 \pm 0$ in that region of the CMD, which will clearly make it the most important part of the fit.  To compensate, a small constant value was added to the $\sigma^2$ denominators of the $\chi^2$ calculations.  Additionally, the red clump, horizontal branch, and AGB regions of the CMD were given less weight in the fits, again so that unfittable parts of the CMD would be less important in the fitting.  Finally, $\chi^2$ value for any bin not allowed to exceed 10, so that no one piece of the CMD would determine the fit.  This modified $\chi^2$ goodness-of-fit parameter was thus used in by an amoeba algorithm to solve for the star formation rates, given the assumed enrichment, distance, and extinction.

In order to make a search for those values as well, a large number of fits was made, each with a different combination of initial metallicity, current metallicity, metallicity enrichment function, distance, and extinction.  All solutions with a modified $\chi^2$ near that of the best fit were averaged together for the determination of best values and uncertainties of the metallicities, distances, extinctions, and star formation rates.  Thus a parameter that strongly affected the solution would give a small uncertainty, while one (such as initial metallicity) with only a weak effect would give a large uncertainty.  Additionally, any relations between parameters (such as the age-metallicity-distance ``degeneracy'') are accounted for in the uncertainties.

\subsection{Era of Initial Star Formation}

In addition to the distance and extinction, the time of initial star formation in WLM is an essential parameter to calculate before determining the galaxy's star formation history.  This is because the evolutionary models reproduce neither the horizontal branch nor the red clump structures well, and thus a blind solution using the models can easily be thrown off in dealing with these old populations.  In the case of WLM, doing so would cause the models to attempt recreating the tight red clump as a red horizontal branch.  But clearly the field has little or no horizontal branch, which sets the upper limit on large amounts of star formation to $\sim$12 Gyr ago.

To determine the age of initial star formation, a set of history calculations was made that were sensitive to the age of initial star formation, with the distance and extinction adopted from above and the enrichment history allowed to vary.  Of the possible starting points, the oldest (12 Gyr) gave the best fits and is used below.  The question of whether the galaxy actually had \textit{no} star formation from the formation of the globular cluster (14.8 $\pm$ 0.7 Gyr ago) to 12 Gyr ago, or whether the galaxy had a non-zero but very small star formation rate cannot be solved from this data, as there are stars where the horizontal branch would be expected, but no horizontal feature is seen in the CMD, placing an upper limit of $\sim2 \times 10^{-5} M_\odot$/yr on the star formation rate between 12 and 15 Gyr ago.  Regardless of whether or not the galaxy waited until 12 Gyr to form any stars, it clearly waited until that point to form a measurable number of stars.

\subsection{Global Star Formation History and Enrichment}

The star formation history study was broken into two parts.  First was the study of the global star formation history and enrichment, which used the combined CMD from all three WF chips.  The distance and extinction were also determined here, permitting a self-consistent solution to be found.  For the size of the region studied ($\sim$500 pc), any stars formed more than a Gyr ago should be mixed by the present time (with a random velocity of 1 km/sec, a 500 pc region would be mixed in 500 Myr), and thus the combination of the CMDs will simply improve the signal-to-noise of the results.  Additionally, the best star formation history information is only available to 1.5 Gyr, beyond which point the main sequence drops below the limit of good photometry, so the signal-to-noise improvement obtained by combining the fields is essential.

To study the global star formation history, a set of star formation history calculations was made allowing metallicity enrichment history (initial metallicity, current metallicity, and function of metallicity increase), distance, and extinction to vary, with the best-fitting solutions averaged with a weighted average to determine the best results.  Uncertainties were calculated by adding the standard deviations from the fitting procedure to the standard deviations of Monte Carlo simulations in quadrature, so that both uncertainties resulting from the parameters and those from the fit itself are included in the final results.  This solution was made using 9335 stars with $20.5 < V < 26.5$ and $20 < I < 26$.

The measured distance modulus was $(m-M)_0 = 24.88 \pm 0.09$, with an extinction of $A_V = 0.04 \pm 0.06$.  This distance is somewhat larger than that measured for the globular cluster ($(m-M)_0 = 24.73 \pm 0.07$ in Hodge et al. 1999, later revised to $(m-M)_0 = 24.77 \pm 0.10$ using improved photometry) and the RGB tip distance measured by Minniti \& Zijlstra (1997) of $24.75 \pm 0.10$, but is within uncertainties of both.  An RGB tip analysis, using the method given by Lee et al. (1993), gives a distance of $(m-M)_0 = 24.90 \pm 0.12$, which is consistent with the above measurement.  This distance is also consistent with the Cepheid distances given in Section 1.

Star formation rates are given assuming a Salpeter IMF with cutoffs at 120 and 0.1 $M_\odot$, which was used out of necessity because this data does not extend beyond the old main sequence turnoff and therefore the IMF cannot be measured.  An IMF error for the old stars would introduce a gradient into the data.  For example, using too shallow of an IMF would generate fewer stars in the older synthetic CMDs, and thus return too large of a star formation rate in the very old times.  The lower cutoff is also a simplification, again made because the data contains no information about low-mass stars.  Most likely the true mass of stars formed is lower than the numbers given here, scaled by some constant factor.

Results are shown in Table \ref{tab-oldsfh} and Figure \ref{fig-oldsfh}.

\placetable{tab-oldsfh}
\placefigure{fig-oldsfh}
\placefigure{fig-synth}

A synthetic CMD for the region reconstructed from the models, artificial star results, and determined star formation history is shown in Figure \ref{fig-synth}a.  A glance shows that the fit appears to be good, with the primary differences between the two diagrams (red clump position, strongly populated blue loop, and brighter AGB) located in the regions where theoretical models are highly uncertain.

The lower part of the visible main sequence is more populated in the reconstructed CMD, which is most likely a result of the assumption of constant star formation rate across the time of each bin.  Since the youngest bin lasts from 200 Myr ago to the present, any change in the star formation rate over that time will alter the MS luminosity function and not be reproduced in the synthetic CMD.  In this case, the recent star formation rate is likely higher than the 200 Myr average, thus producing a flatter MS luminosity function in the observed data.  At any rate, the recent star formation rates are studied in the next section, and will not be dealt with here.

It is encouraging that the CMD features that are expected to be well-reconstructed, namely the positions of the main sequence and the red giant branch, are indeed correct.  The reconstructed RGB is somewhat too broad, a result of the uncertainties in modeling old star formation rates and metallicity enrichment.

It should be stressed that there were no built-in assumptions of the metallicity, distance, or extinction in this star formation history solution; the star formation history program was able to determine all of these values and produce a synthetic CMD that closely resembles the observed CMD.

\placefigure{fig-comp}
\placefigure{fig-comp2}

A more detailed comparison is shown in Figures \ref{fig-comp} and  \ref{fig-comp2}, which show the binned CMDs that were used for the solution (0.1 magnitude resolution in $V-I$ and 0.3 magnitude resolution in $V$ for Figure \ref{fig-comp}; 0.05 $V-I$ and 0.15 $V$ for Figure \ref{fig-comp2}), as well as the residuals and the fit parameter values.  Magnitude limits in these plots are $20 < V < 26$ and $-0.5 < V-I < 2.5$.  The subtracted CMDs are shown at 2.5 times the contrast level of the two CMDs, while the modified $\chi$ plots are shown on a scale from -9 to +9 $\sigma$.  For both, a point darker than average is one with more synthetic points than observed.

The largest errors in fitting the CMD come in the red clump and horizontal branch regions, which along with the AGB are the regions with the highest uncertainty in the models.  Thus errors here are not surprising, and do not necessarily mean that the fit quality is poor.  Most notably, the models still show a slight red RGB, despite the 12 Gyr age limit placed on the fit, while the observed CMD shows a red clump slightly brighter and bluer.
Errors from the red clump region are up to 9 $\sigma$ (in the undersubtracted region bluewards of the red clump and blue loop base).  Additionally, the theoretical AGB has a much different slope than that observed here, thus making it impossible to adequately model.  However, the erroneous AGB modeling provides very little to the fit (no more than a 1 $\sigma$ error at any point), as it contains few stars.

The red giant branch is another source of error, with the red edge fit very well, but the blue edge problematic.  The blue edge is drawn out by the program's attempt to fit the bluest red giant branch stars as the old population.  A solution which breaks the old bin (9-12 Gyr old) into multiple bins would solve this problem, allowing a more flexible enrichment and star formation history, although this would imply better resolution than is really possible in this type of work.
Undersubtraction of the observed blue edge of the RGB produces errors of up to 6 $\sigma$, while oversubtraction in the modeled blue edge produces errors of up to 3 $\sigma$.

As noted above, the main sequence is oversubtracted at the faint end ($V > 24$).  This contributes a 3 $\sigma$ error in the solution, and is the result of changes in the star formation rate within the past 200 Myr.

Looking at the calculated star formation history, it seems that the star formation of WLM began with a strong initial star formation episode, beginning approximately 12 Gyr ago.  This event contained about half (by mass) of the total star formation that WLM has had over its lifetime.  The star formation rate fell considerably after that episode, although uncertainties are sufficiently large that nothing can be said for certain about the period between 2.5 and 9 Gyr ago except that there was some star formation, and likely with an average rate of 1 or 2 $\times 10^{-4} M_\odot$/yr.  The recent star formation history will be covered in the next section.

The metallicity enrichment of WLM shows an initial abundance of -2.18 $\pm$ 0.28, although a glance at the CMD tells that few stars formed with such a low metallicity.  The metallicity at the tail end of the 3 Gyr initial star-forming episode was -1.34 $\pm$ 0.14.  This result is consistent with the result of Minniti \& Zijlstra (1997), that the average metallicity of old RGB stars is -1.45 $\pm$ 0.2, as well as the Hodge et al. (1999) measurement of the globular cluster metallicity of -1.51 $\pm$ 0.09 (revised to -1.63 $\pm$ 0.14 with improved photometry).  The metallicity has continually climbed since that time, reaching a current value of -1.08 $\pm$ 0.18.  This value for the current metallicity agrees well with the -1.15$\pm$0.2 found by Hodge \& Miller (1995) for two HII regions in WLM.

\subsection{Recent Star Formation History}

In contrast to the well-mixed old stars, Minniti \& Zijlstra (1997) and Ferraro et al. (1989) found a stellar population gradient in WLM, with the bar region contributing more recent star formation.  As their data extended to $V=\sim24$, it was impossible to quantify this result, but with the HST data here a star formation history analysis can be done separately on different parts of the galaxy.  For the study of recent star formation, the three fields were kept separate, with 6381 stars in WF2, 2165 in WF3, and 789 in WF4.
The analysis was carried out using the distance, extinction, and enrichment calculated above, with results shown in Table \ref{tab-newsfh} and Figure \ref{fig-newsfh}.

\placetable{tab-newsfh}
\placefigure{fig-newsfh}

Reconstructed CMDs of the three fields are shown in Figures \ref{fig-synth}b-d.  As with the combined CMD, the chief differences between observed and reconstructed CMDs are in the blue edge of the RGB, the red clump sharpness and position, and the AGB position, resulting from the models' uncertainties in late phases of evolution.  The reconstructed WF2 field, in particular, shows a very wide red clump, apparently due to overestimated incompleteness and crowding errors in that field.  Otherwise, the fits are excellent, as with the combined CMD.

As shown by the global star formation history, WLM has had a burst of star formation activity during the past Gyr and possibly longer.  Determining star formation rates beyond 2.5 Gyr is much more difficult, as the main sequence is below the limit of the good photometry, so only star formation rates to 2.5 Gyr are shown for the fields.  The WF2 field, which is the closest to the galaxy's center (and includes the outer regions of HII complexes C1 and C2 as well as including all of HII region 1 listed in Hodge \& Miller 1995), contains most of the recent star formation (85\% of the star formation in the past 200 Myr), while the 1000 to 2500 Myr star formation rate is more evenly distributed between the regions, which can be explained through mixing.

The Hodge \& Miller (1995) data on WLM provide an additional constraint on the current star formation, as they find a formation timescale (defined as total star formation amount divided by current star formation rate) of 150 Gyr for the galaxy, significantly longer than the value of 14 Gyr that would be calculated from the current star formation rate shown in Table \ref{tab-oldsfh}.  (Interestingly, the 14 Gyr calculated from the data presented here implies that the galaxy's current star formation rate is roughly equal to the average rate for its lifetime.)  Two possible causes of this difference can be pinpointed.  First, our sample consists of only a small fraction of WLM (one including a small HII region and bordering two large HII complexes), and the global star formation timescale could be significantly different from that found in this region.  Second, it is possible that the estimate of the lifetime-averaged star formation rate presented here is more accurate, as this is the first star-formation history work for WLM using HST WFPC2 data.  However, without data of HST quality elsewhere in the galaxy, it is likely not possible to determine which (or perhaps a combination of the two) is the source of the difference.

Overall, the results confirm the previous findings of Minniti \& Zijlstra (1997) and Ferraro et al. (1989).  The WF2 field, which is mostly in the WLM bar, has seen the most star formation activity in the recent burst of activity, accounting for $\sim$85\% of the total recent star formation.  (For comparison, the WF2 field contains just 62\% of the stars with $V-I > 0.5$.)  The WF3 field, which is located on the edge between the bar and the halo, is intermediate, containing 12\% of the recent star formation and 26\% of the red stars.  Finally, the WF4 field, located entirely in the halo, appears to have almost no recent star formation (3\% of the total) despite containing 11\% of the red stars.

\section{Summary}

HST F555W and F814W photometry of a portion of the WLM galaxy has been presented.  A distance modulus of 24.88 $\pm$ 0.09, is calculated using CMD analysis and shown to be consistent with a distance determination from the RGB tip and previous measurements by Hodge et al. (1999), Minniti \& Zijlstra (1997), Sandage \& Carlson (1985), and later Cepheid recalibrations.  The galaxy's oldest stars appear to be $\sim$12 Gyr old, with an abundance [Fe/H] of -2.18 $\pm$ 0.28 for the initial star-forming episode.  It is calculated that roughly half of the galaxy's star formation occurred before 9 Gyr ago.

Star formation rates in the intermediate ages are more difficult to calculate, because of no main sequence stars and poor modeling of AGB stars by the evolutionary tracks used.  The metallicity has risen to a current value of [Fe/H] = -1.08 $\pm$ 0.18, consistent with the Hodge \& Miller (1995) measurement of HII region abundances in WLM.

Increased activity has taken place recently, for approximately the past 1 Gyr.  The activity for stars older than this cannot be pinpointed because of mixing within the galaxy, but the current star formation is limited to the bar of the galaxy.  This activity has continued to the present, with a measured formation timescale of 14 Gyr given the current star formation rate.

\acknowledgments

I am indebted to the staff of the Space Telescope Science
Institute for obtaining these data and to NASA for support of the analysis
through grant GO-06813.

\clearpage

\figcaption[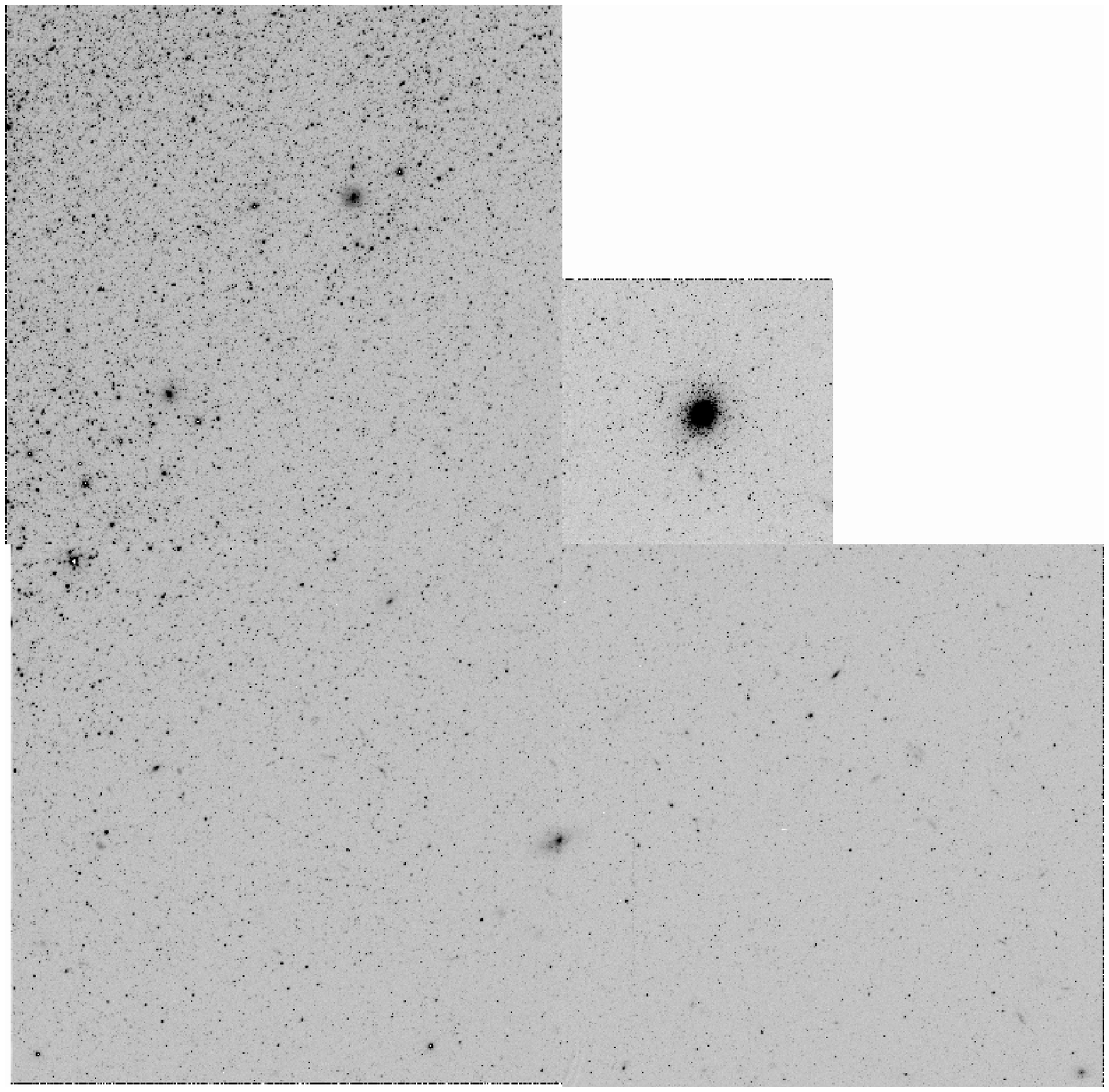]{Combined $V$ image of WLM after cosmic ray cleaning.  WF2 is in lower left, WF3 in lower right, and WF4 in upper right. \label{fig-image}}
\figcaption[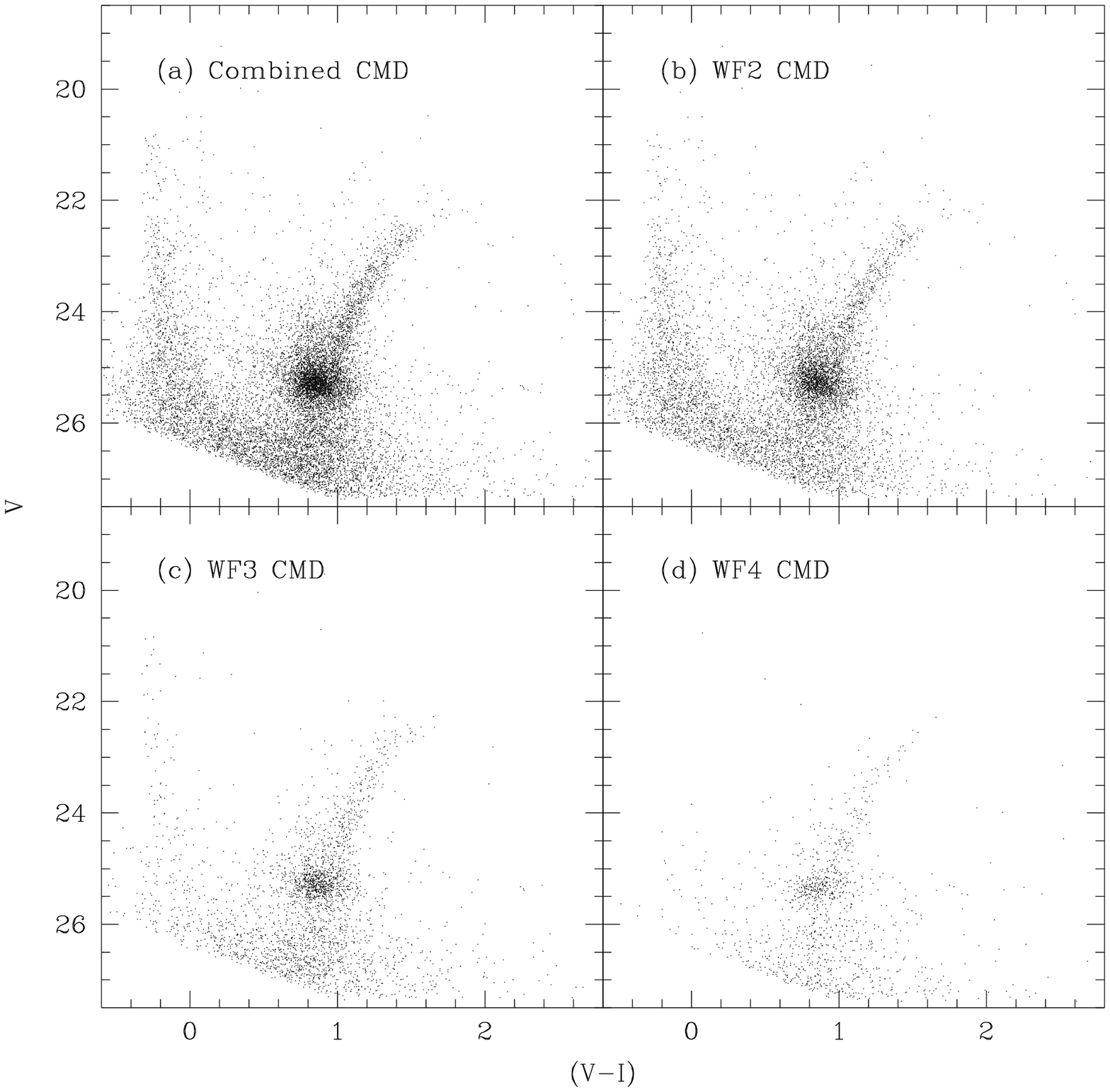]{Color-Magnitude Diagrams of WLM \label{fig-cmds}}
\figcaption[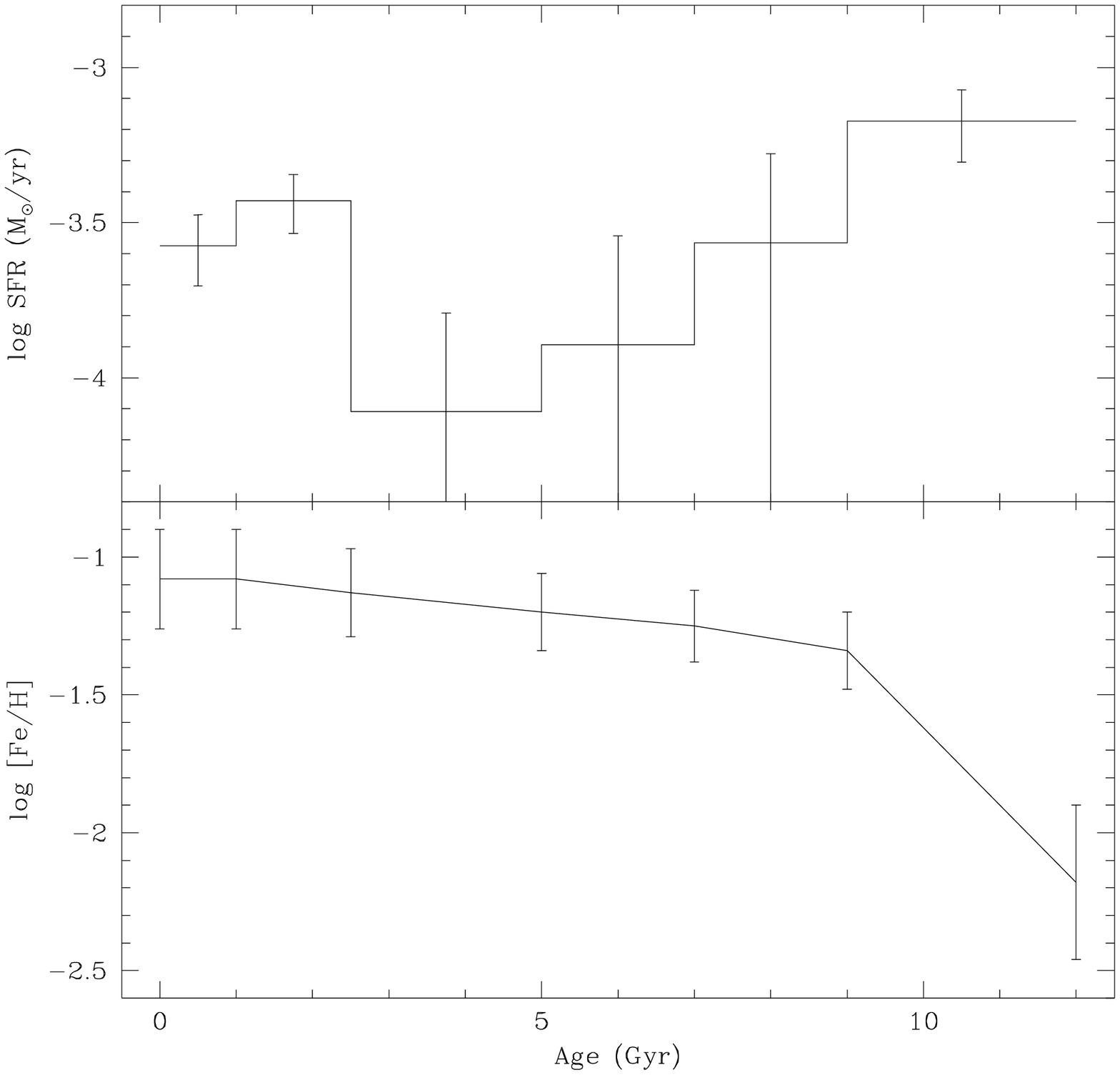]{Star Formation History of WLM \label{fig-oldsfh}}
\figcaption[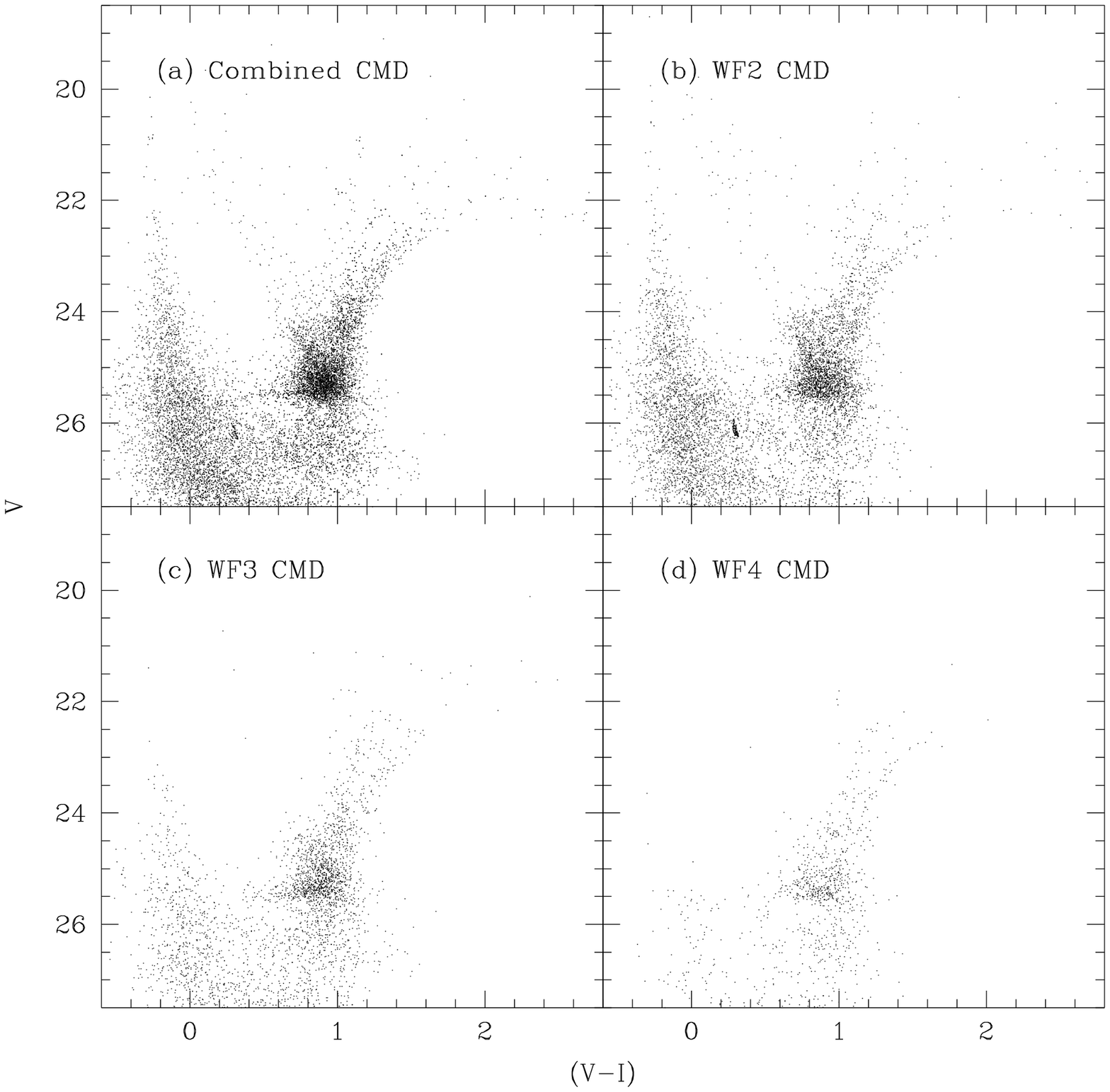]{Reconstructed Color-Magnitude Diagrams of WLM \label{fig-synth}}
\figcaption[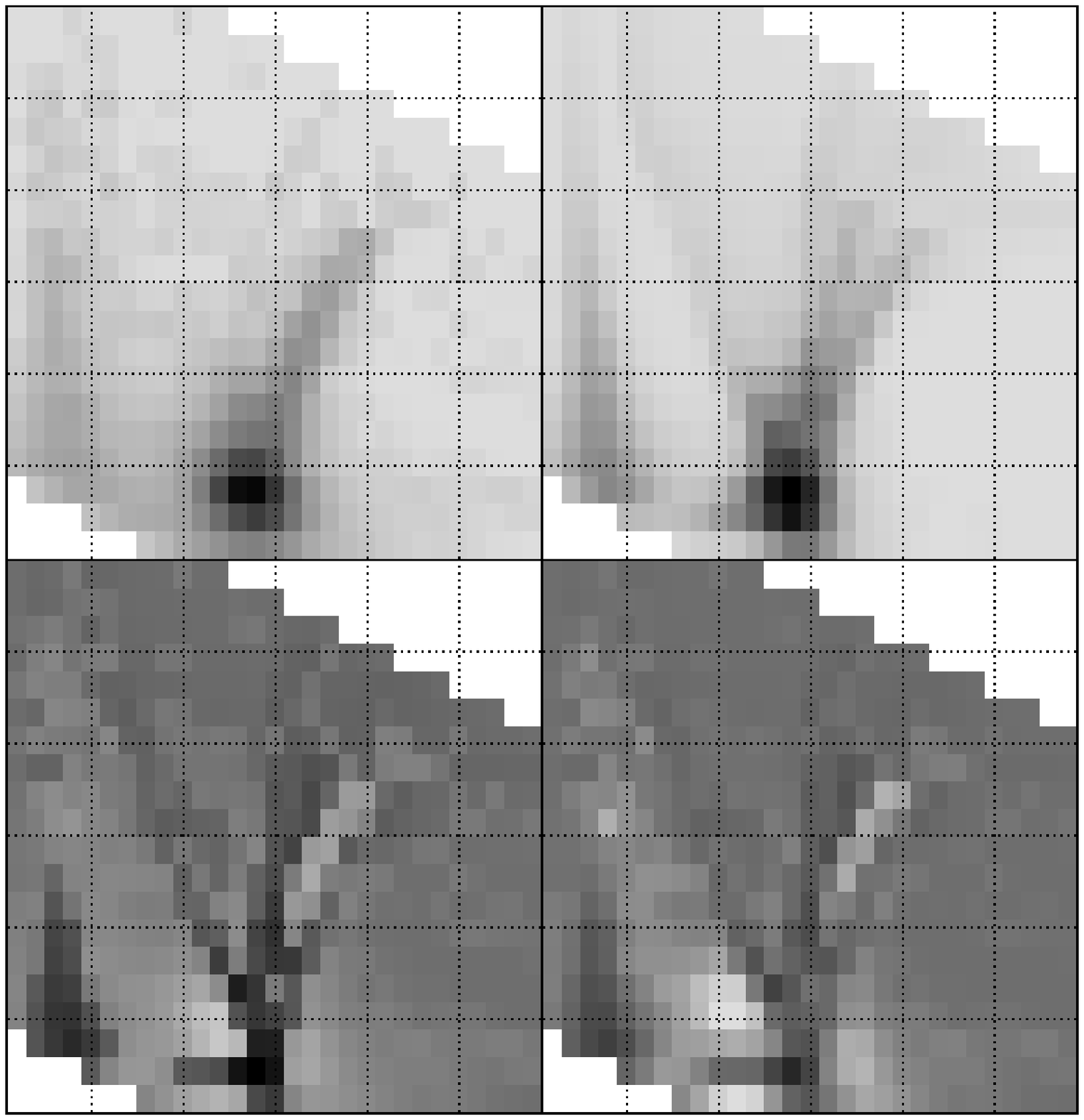]{Comparison of Observed and Reconstructed CMDs.  Upper left is the binned observed CMD; upper right is the binned reconstructed CMD.  The lower left shows the subtracted diagram, while the lower right shows the modified $\chi$ values at each location.  See text for explanation. \label{fig-comp}}
\figcaption[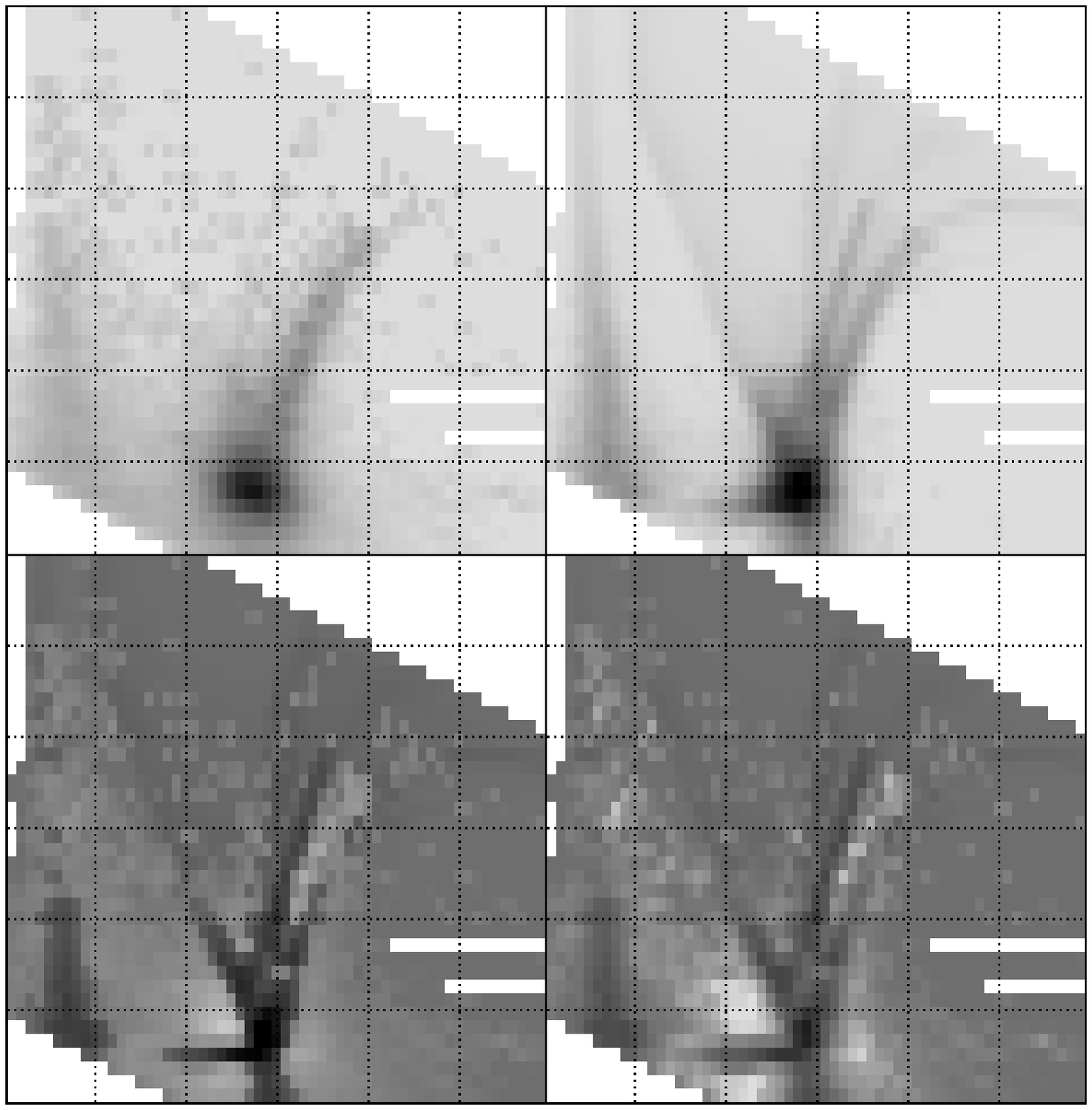]{Comparison of Observed and Reconstructed CMDs.  Same as Figure \ref{fig-comp} \label{fig-comp2}}
\figcaption[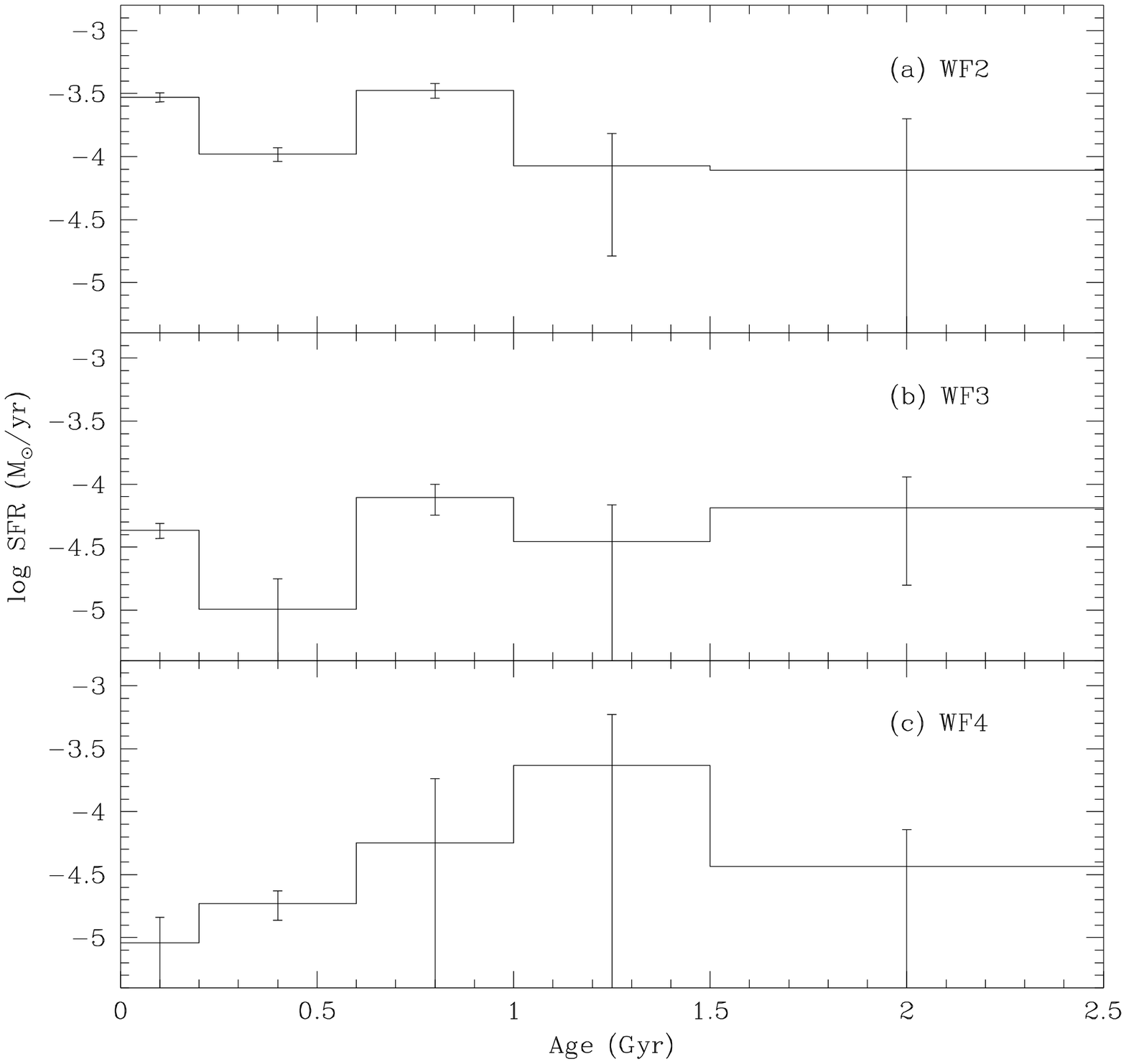]{Recent Star Formation History of WLM Fields \label{fig-newsfh}}

\clearpage
\begin{deluxetable}{crr}
\tablenum{1}
\tablewidth{0pt}
\tablecaption{Global Star Formation History of WLM \label{tab-oldsfh}}
\tablehead{
\colhead{Age (Gyr)} &
\colhead{SFR ($10^{-5} M_\odot$/yr)} &
\colhead{[Fe/H]$_0$}}
\startdata
 0$-$1   &  27 $\pm$  6 & -1.08 $\pm$ 0.18 \nl
 1$-$2.5 &  37 $\pm$  8 & -1.13 $\pm$ 0.16 \nl
 2.5$-$5 &   8 $\pm$  8 & -1.20 $\pm$ 0.14 \nl
 5$-$7   &  13 $\pm$ 15 & -1.25 $\pm$ 0.13 \nl
 7$-$9   &  27 $\pm$ 25 & -1.34 $\pm$ 0.14 \nl
 9$-$12  &  67 $\pm$ 15 & -2.18 $\pm$ 0.28 \nl
\enddata
\end{deluxetable}

\clearpage
\begin{deluxetable}{crrr}
\tablenum{2}
\tablewidth{0pt}
\tablecaption{Recent Star Formation History of WLM Fields \label{tab-newsfh}}
\tablehead{
\colhead{Age (Gyr)} &
\colhead{WF2 SFR ($10^{-5} M_\odot$/yr)} &
\colhead{WF3 SFR ($10^{-5} M_\odot$/yr)} &
\colhead{WF4 SFR ($10^{-5} M_\odot$/yr)}}
\startdata
0.0$-$0.2    & 30 $\pm$  3 &  4 $\pm$ 1 &  1 $\pm$  1 \nl
0.2$-$0.6    & 10 $\pm$  1 &  1 $\pm$ 1 &  2 $\pm$  1 \nl
0.6$-$1.0    & 34 $\pm$  4 &  8 $\pm$ 2 &  6 $\pm$ 13 \nl
1.0$-$1.5    &  8 $\pm$  7 &  3 $\pm$ 3 & 23 $\pm$ 36 \nl
1.5$-$2.5    &  8 $\pm$ 12 &  7 $\pm$ 5 &  4 $\pm$  4 \nl
\enddata
\end{deluxetable}

\end{document}